\documentclass[12pt]{elsart}
\usepackage{graphicx,times,natbib,epsfig}
\begin{document}
\begin{frontmatter}
\title{Extracting first science measurements from the southern
detector of the Pierre Auger observatory}
\author {L. Wiencke for the Pierre Auger Collaboration}
\address[Utah] {Department of Physics, 
University of Utah, Salt Lake City, UT 84112, USA\\ email: wiencke@cosmic.utah.edu}
\begin{abstract}
The world's largest cosmic-ray detector is nearing completion in the
remote Pampas of Argentina. This instrument measures extensive
air-showers with energies from ${10^{18}-10^{20}}$ eV and beyond. A
surface detector array of area 3000 ${km^2}$ records the
lateral distribution of charged particles at ground level. A
fluorescence detector overlooking the surface detector records the
longitudinal light profiles of showers in the atmosphere to make a
calorimetric energy measurement. A ``test beam'' for the
fluorescence detector is generated by a calibrated laser near the
array center. This talk will focus on detector characterizations
essential to the first science results that have been reported from
the observatory. Plans to construct a larger instrument in the
northern hemisphere will also be outlined.
\end{abstract}
\end{frontmatter}
{\bf 1. Introduction}\\ The origin of cosmic rays with energies beyond
$10^{18}$ eV is a long standing puzzle. For each decade of energy
increase the differential flux falls by three orders of magnitude to
approach, at $10^{20}$ eV, a scant one event per $km^{2}$ per century.
This alone hints at the formidable instrumental
challenges. Fore-runner experiments have operated different types of
detectors in different locations. This complicates the understanding
of these data in combination because the differing systematics that
necessarily underlay astrophysical interpretations can not be compared
directly.

\begin{figure}
 \centering
 \includegraphics[width=0.80\textwidth]{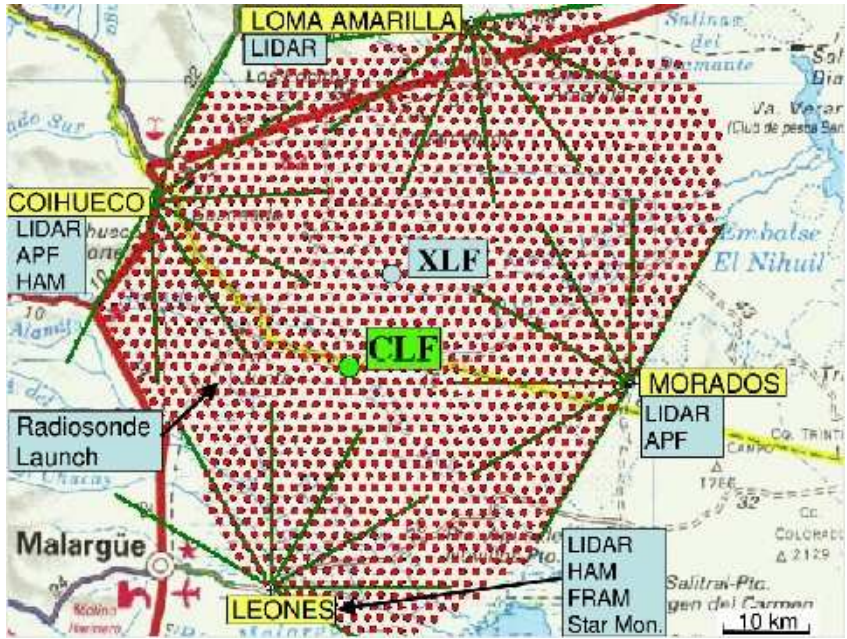}
 \caption{Configuration of the Pierre Auger Observatory southern detector.(see text.)}
 \label{map}
\end{figure}

The Pierre Auger Observatory combines the surface and air-fluorescence
techniques with two co-located detectors. Deployment of the baseline
configuration in the south (Fig. \ref{map}) is about 75\%
complete (1100+ of 1600 tanks, 3 of 4 fluorescence ``eyes'', 3 of
4 LIDARs, 1 of 2 central lasers) and is expected to finish in
2007.

\begin{figure}
\begin{minipage}[t]{7.0cm}
 \includegraphics[width=1.00\textwidth]{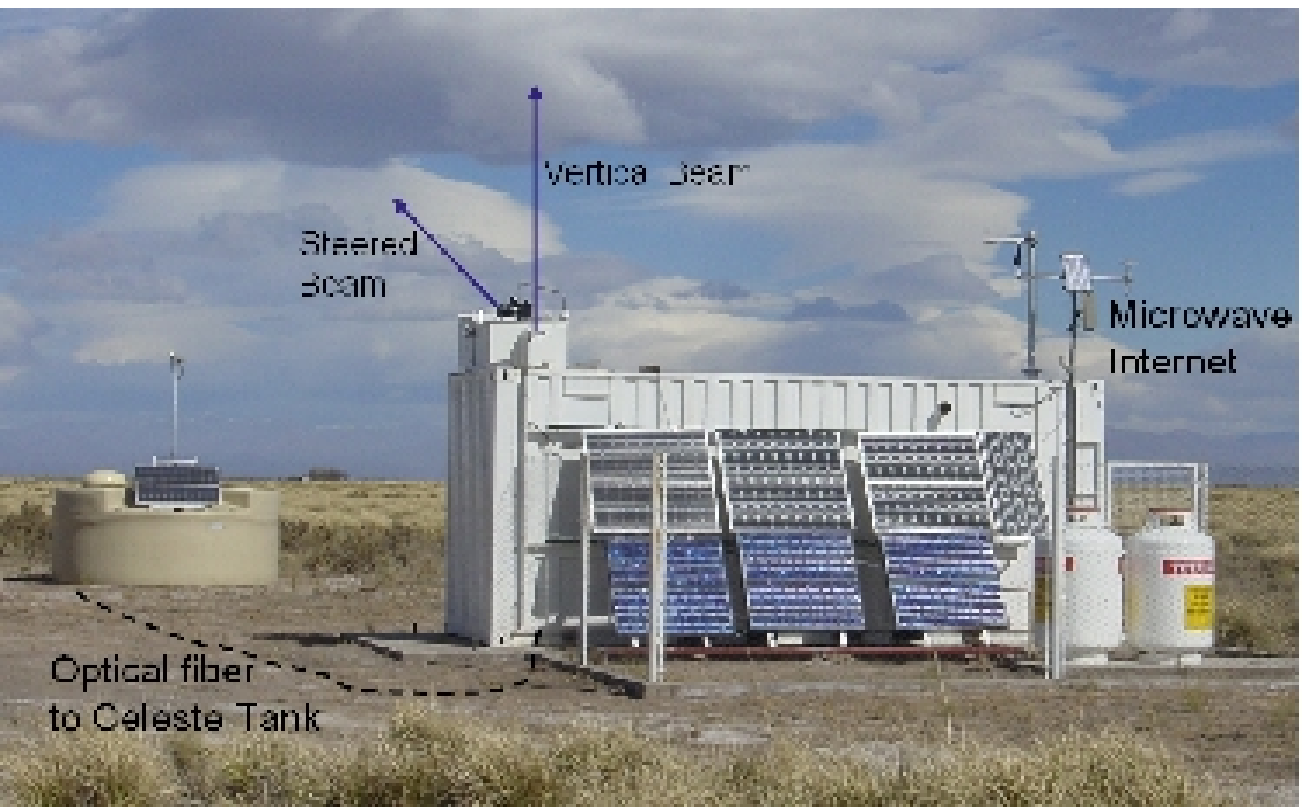}
 \caption{Central laser facility (CLF) and an adjacent surface detector tank.}
 \label{clf}
\end{minipage}
\hfill
\begin{minipage}[t]{6.0cm}
 \includegraphics[width=1.00\textwidth]{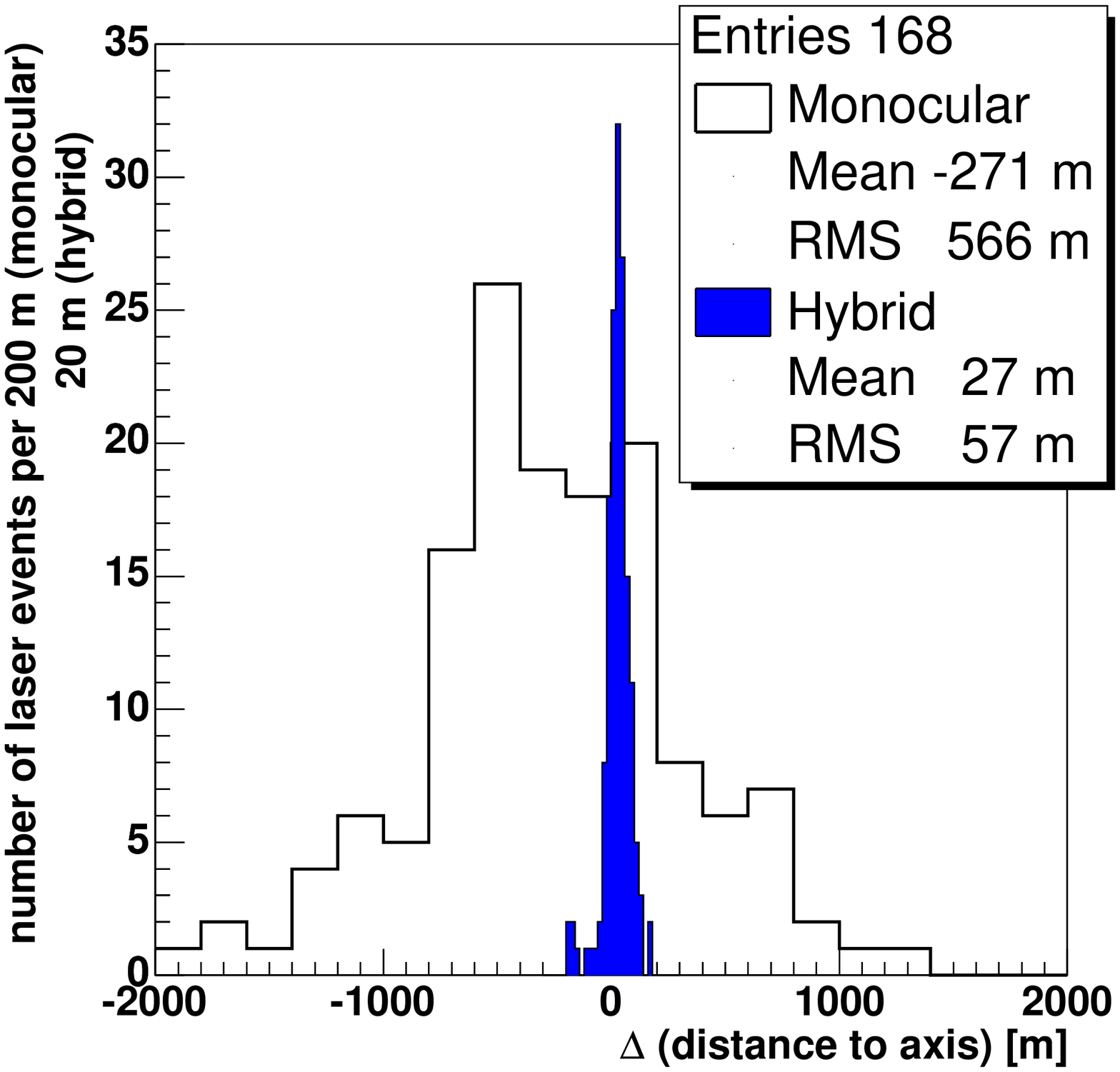}
 \caption{Rp resolution of CLF vertical tracks as reconstructed using mono (1 FD eye
 only) and hybrid (1 FD eye + SD timing) data.}
 \label{clf-rp}
\end{minipage}
\end{figure}

{\bf 2. Pierre Auger Detectors and their Characterization}\\ The
surface detector (SD) \cite{XB2005} of 1600 water cherenkov tanks
samples the charged particle lateral distribution of extensive
air-showers (EASs) at ground level. With continuous operation and 
area of 3000 $km^{2}$, the SD provides the
statistical engine. After correcting for zenith angle, the density of
charged particles 1000 m from the shower core provides an energy
estimator. For a $10^{19}$ eV EAS with zenith angle less than 50
degrees, the statistical uncertainty in S1000 is of order 10\%
\cite{PER2005}. Angular accuracy is a function of the number of tanks
triggered, which depends primarily on shower energy and zenith
angle. For EASs measured by at least 5 tanks, this accuracy is better
than 1 degree \cite{CB2005}. Essentially all events above $10^{19}eV$
meet these criteria.

A fluorescence detector (FD) of four ``eyes'' overlooking the SD
operates at night to record the longitudinal light profiles of EASs as
they develop through the atmosphere. This calorimetric energy
measurement does not rely on interaction models; the amount of
fluorescence light emitted is proportional to the energy deposited in
the atmosphere. At present the systematic energy uncertainty is less
than 25\% (Table \ref{errors}). This value is expected to decrease as
improvements are expected in the three largest terms: fluorescence
yield, detector photometric calibration, and atmospherics. About 10\%
of the observed EASs are recorded by both detectors (FD and SD) to
construct a ``Rosetta stone'' hybrid data sample now in excess of
30,000 events.

\begin{figure}
\begin{minipage}[t]{6.0cm}
 \includegraphics[width=1.00\textwidth]{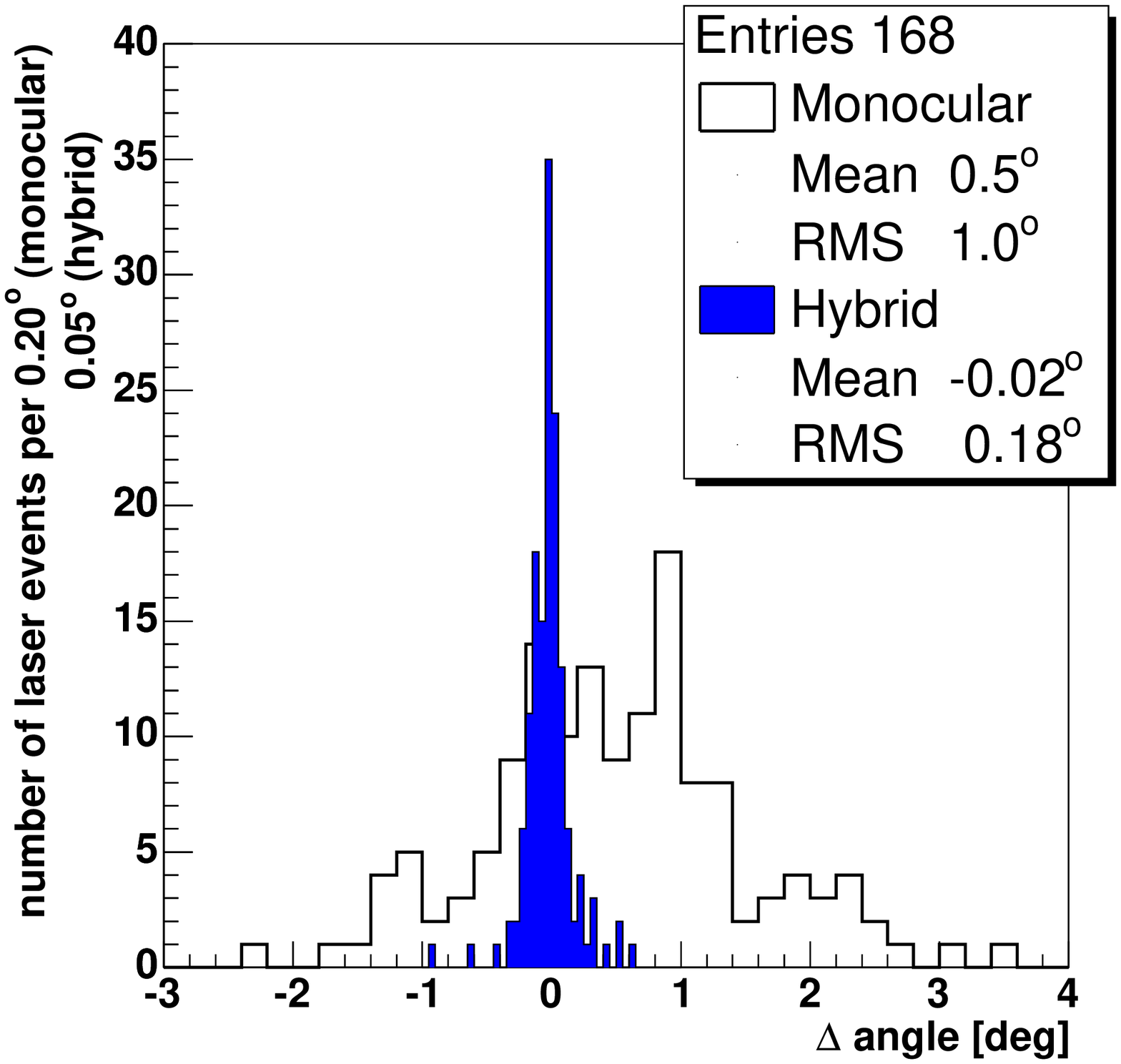}
 \caption{Angular resolution of CLF vertical tracks as reconstructed using mono (1 FD eye
 only) and hybrid (1 FD eye + SD timing) data.}
 \label{clf-ang}
\end{minipage}
\hfill
\begin{minipage}[t]{7.0cm}s
 \includegraphics[width=1.0\textwidth]{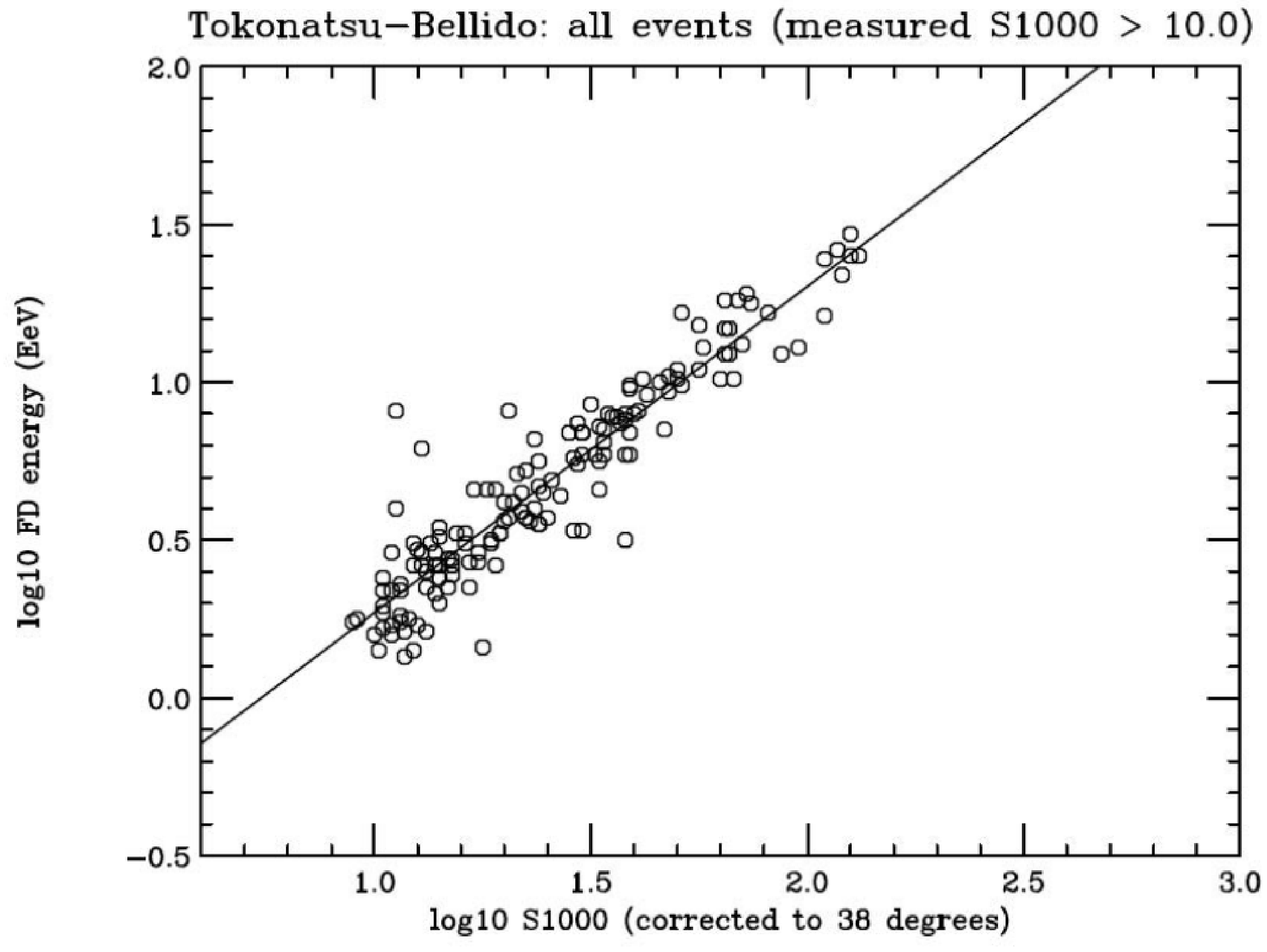}
 \caption{Energy as measured by the FD (vertical axis) vs s1000 as measured by
 the SD (horizontal axis).}
 \label{e38new}
\end{minipage}
\end{figure}

One important role of the FD, through hybrid data, is to verify the
pointing accuracy and to determine the energy scale of
the SD \cite{S2005}. To realize this, the Auger design incorporates
various methods and instruments to determine the geometrical reconstruction
accuracy\cite{M2005} and the photometric energy scale of the
FD\cite{PB2005} and to measure the density profile \cite{K2005} and
the distribution of aerosol in the atmosphere through which the FD
observes EASs \cite{R2005}.

Test beams of $E>10^{17} eV$ particles are unavailable. Pulsed UV
lasers, however, provide a practical alternative. The total amount of
light scattered from a 5 mJ energy UV beam pulse directed vertically
into the atmosphere is roughly equivalent to the amount of
fluorescence light emitted by a shower with energy in the region of
the predicted GZK suppression. The baseline configuration includes two
laser facilities, dubbed the CLF \cite{A2005} (Fig. \ref{clf}) and the
XLF (under construction), near the middle of the SD. The remotely
controlled CLF operates during FD data collection to send light
simultaneously via optical fiber into a SD tank and via mirrors into
the sky. The hybrid laser data sample is used to measure the relative
timing between the FD and the SD\cite{AL2005}, and to realize
significant improvements in geometric reconstruction by invoking the
constraint of SD timing (Figs. \ref{clf-rp},\ref{clf-ang}). Fitting
the longitudinal profile of CLF tracks observed by the FD provides an
hourly data base of aerosol optical depth measurements. A comparison
between the laser pulse energy and the energy as reconstructed from FD
measurements tests, under aerosol-free conditions, the photometric
calibration of the FD and the modeling of molecular atmosphere
over a total light path of 60 km. The current level of
agreement is better than 15\%.

Analysis of hybrid EAS data led to many of the first science
results. Hybrid events have set the energy scale (Fig. \ref{e38new})
for the first energy spectrum measurement \cite{S2005}, set a limit on
the flux of neutral EeV particles pointing back to the galactic center
\cite{AS2005}, and set limits on the photon fraction of shower
composition \cite{RI2005}.
\begin{table}
 \centering
\begin{tabular}{rlrl}
\hline
Term & Error(\%) & Term & Error(\%)\\
\hline
Light collection & 5 & Atmosphere (aerosols) & 10\\
Detector photometric calibration & 12 &Atmosphere (clouds) & 5\\
Geometric reconstruction & 2 & Atmosphere (density profile) & 2\\
Correction for Missing Energy & 3 & Fluorescence yield & 15\\
\hline
& {\bf Quadrature Sum = 23} & & \\
\hline\\
\end{tabular}
\caption{Current estimates of systematic error components to the FD energy measurement.}
\label{errors}
\end{table}
\\
\\{\bf 3. The Northern Detector and Enhancements in the South}\\ The
collaboration has selected southeast Colorado to site the northern
detector. Much of this area is accessed by a mile square grid of roads
and could support deployment of an array much larger than that in the
south. A proposal is in preparation. Enhancement of the southern
detector is also anticipated by adding muon counters, in-fill
surface detectors and higher elevation angle FD telescopes to reduce
the energy threshold towards $10^{17} eV$. R\&D in radio detection is
also planned.

\end{document}